\documentclass[letterpaper]{article}
\usepackage{natbib}
\usepackage{alifexi}
\usepackage{epsfig}
\usepackage{subfigure}
\title{The Efficacy of Group Selection is Increased by Coexistence Dynamics within Groups}
\author {Simon T. Powers, Alexandra S. Penn \and Richard A. Watson\\
\mbox{} \\
       School of Electronics \& Computer Science, University of Southampton, Highfield, Southampton SO17 1BJ, UK \\
       Simon.Powers@Unil.ch}

\begin{document}
       \maketitle

\begin{abstract}

Selection on the level of loosely associated groups has been suggested as a route towards the evolution of cooperation between individuals and the subsequent formation of higher-level biological entities. Such group selection explanations remain problematic, however, due to the narrow range of parameters under which they can overturn within-group selection that favours selfish behaviour. In principle, individual selection could act on such parameters so as to strengthen the force of between-group selection and hence increase cooperation and individual fitness, as illustrated in our previous work. However, such a process cannot operate in parameter regions where group selection effects are totally absent, since there would be no selective gradient to follow. One key parameter, which when increased often rapidly causes group selection effects to tend to zero, is initial group size, for when groups are formed randomly then even moderately sized groups lack significant variance in their composition. However, the consequent restriction of any group selection effect to small sized groups is derived from models that assume selfish types will competitively exclude their more cooperative counterparts at within-group equilibrium. In such cases, diversity in the migrant pool can tend to zero and accordingly variance in group composition cannot be generated. In contrast, we show that if within-group dynamics lead to a stable coexistence of selfish and cooperative types, then the range of group sizes showing some effect of group selection is much larger. 
  
\end{abstract}

\section{Introduction}

The evolution of cooperation between biological individuals, both generally and as a vital part of the formation of new higher-level composite individuals, is an important and much discussed open question in both evolutionary biology \citep{Smith:1995:a, Keller:1999:a, Michod:1999:a, Hammerstein:2003:a, Okasha:2006:a} and artificial life \citep{Bedau:2000:a}. The fundamental problem is that any group of cooperative types, whose members donate some component of individual fitness in order to benefit their group, is vulnerable to invasion by selfish cheats that reap the group benefits of cooperation without paying the individual cost. Various mechanisms by which cooperation can nevertheless evolve have been suggested, including so-called `group selection' models in which population structure exists such that individuals spend part of their time in groups (rather than freely mixed in the whole population) before mixing in a migrant pool from which new groups are formed \citep{Wilson:1980:a}.  Although selection within any given group will always favour selfish individuals, groups with a higher proportion of cooperators are more productive and hence contribute more individuals to the migrant pool and the next generation of groups. 
 
Such models have been found to allow cooperation to evolve or be preserved in a population, but only within certain narrow parameter ranges (discussed below). This clearly presents a problem when appealing to such mechanisms as a route towards the evolution of higher-level individuals. As we have argued in previous work \citep{Powers:2007:b}, the conditions that allow group selection to be effective and control its strength need not be externally imposed. Instead, they may be products of individual characters and hence subject to individual selection. Specifically, if key parameters such as group size are subject to individual adaptation (for example, via production of extracellular matrix in a bacterial biofilm), then a process akin to niche construction \citep{Smee:2003:a} supporting the evolution of cooperation may occur whereby cooperative traits and those affecting the strength of group selection evolve concurrently.

However, this process would require the existence of an adaptive gradient, such that a small parameter change could increase the strength of group selection and consequently cooperation and individual fitness. In this paper we consider one important parameter, group size. Where groups are formed randomly, existing models have shown that increasing group size rapidly causes the measurable effect of group selection to reach zero \citep{Wade:1978:a}. This then means that for a large region of parameter space, a small decrease in group size, e.g. by individual mutation, would have no effect on fitness, i.e. there would be no selective gradient towards smaller groups and increased individual fitness. However, the model developed and presented in this paper suggests that the rapid tendency of group selection effects to zero is a consequence of an assumption of directional within-group selection. Specifically, classical group selection models attempt to explain the global promotion of a cooperative allele that is driven extinct at within-group equilibrium through directional selection for a rival selfish allele. While this assumption of directional selection and hence competitive exclusion of types is commonplace in population genetics models, a competitive \emph{coexistence} of two types is instead often permitted in ecological models. For example, the classical Lotka-Volterra competition equations allow for coexistence as well as exclusion \citep{May:1976:a}. Such dynamics are of relevance to potential multi-species group selection scenarios; for example, during egalitarian major evolutionary transitions in which different unrelated individuals eventually form a new level of selection, and within extant multi-species consortia such as bacterial biofilms \citep{Burmolle:2006:a} in which group selection effects may be pertinent. \footnote{Coexistence dynamics may also apply to single-species scenarios under certain regimes such as balancing selection (see below) and are hence relevant to group selection discussions more generally.} Results presented in this paper show that where a within-group stable coexistence of types exists, measurable group selection effects are sustained over a much larger range of group sizes, potentially providing an individual adaptive gradient towards smaller groups that enhance group selection over a much larger range of parameter space.

\section{The Limits of Group Selection}

Group selection theory is typically understood as the idea that the differential productivity of groups can lead to changes in allele frequency in a gene pool, in a manner analogous to individual selection acting through the differential productivity of individuals \citep{Wade:1978:a, Wilson:1980:a}. In this paper we are concerned with type 1 group selection, which defines group fitness as the average fitness of the group members \footnote{Type 2 defines group fitness as number of offspring \emph{groups}.} \citep[chap. 2]{Okasha:2006:a}. Models of this process therefore seek to investigate the effect of group structure on the evolution of individual social traits such as cooperation. In particular, a trait that is individually disadvantageous may nevertheless evolve if it has a positive effect on the group as a whole, an idea that dates back to \citet{Darwin:1871:a}. Using the language of multilevel selection theory, such cooperative traits are selected against within-group, since they confer a fitness disadvantage relative to other group members, but are favoured under between-group selection, since they increase group productivity, i.e. average absolute group member fitness. Whether or not the trait spreads in the global gene pool then depends on the balance of these two selective forces.

Three factors are pertinent in determining the outcome of such a scenario. The first is the individual cost to group benefit ratio of the cooperative act; a large individual cost strengthens within-group selection against cooperation, while a large group benefit strengthens between-group selection. Secondly, a group mixing mechanism must exist so that the increased productivity of more cooperative groups may affect the global allele frequencies. We consider here a multi-generational variant of D.S. Wilson's \citeyearpar{Wilson:1980:a} trait-group model, where a global mixing stage occurs every $t$ generations in which the progeny of all groups disperse, join a global migrant pool, and then reform new groups of random composition. Since cooperative groups will contain more individuals prior to dispersal, they will constitute a larger fraction of the global migrant pool, thereby biasing the global allele frequencies. How often global mixing occurs is therefore a crucial parameter, since groups must be mixed before the within-group equilibrium has been reached, otherwise there will be no difference in group productivity for selection to act on (assuming selection within all groups leads to the same group equilibrium \citep{Wilson:1992:a}).

The third factor, and the one which is most often commented on in the group selection literature, is the requirement for there to be variation in the groups' allelic composition. The larger this variance, the greater the difference in group productivity (since within-group dynamics are assumed to be deterministic) and likewise the effect of group selection. Many theoretical models have shown that if group composition is random then very small initial group sizes are needed in order to produce the between-group variance necessary for a measurable effect (see \citep{Wade:1978:a} for a classical review). It is therefore usually concluded that some kind of non-random group formation is required in order for group selection to have any significant effect. The most common way that such assortative grouping is believed to occur in nature is through kin grouping, where the group members are related by descent from a common ancestor and are hence more similar to each other than to members of other groups. Consequently, kin selection \citep{Hamilton:1964:a, Hamilton:1964:b} is commonly seen as the only pertinent force in social evolution. However, even with only simple random sampling, the range of group sizes which produce a group selection effect may be strongly affected if the assumption of within-group competitive exclusion is relaxed. This is because possible between-group variance from sampling error is controlled by the frequency of the least frequent type in the migrant pool. In the competitive exclusion case the possible variance rapidly tends to zero as the selfish type approaches fixation. If, however, coexistence of the two types is possible at equilibrium then we would expect the possible variance to be maintained above zero even at within-group equilibrium frequencies.

\subsection{Group Competition: Aggregation and Dispersal Through a Migrant Pool}

In this paper, we consider a model of a population structure where individuals reproduce in groups for a number of generations. After this period of reproduction within groups, all groups disperse and their progeny mix freely together in a migrant pool. Thereafter, new groups are created from the individuals in the migrant pool, and the process repeats. This process is a multi-generational variant of the trait-group model of \citet{Wilson:1980:a,Wilson:1987:a}, and corresponds closely to the ``Haystack'' model of \citet{Smith:1964:a}. Examples of natural populations that fit this model particularly well include soil-dwelling populations of micro-organisms that are occasionally mixed together during rainstorms \citep[p. 22]{Wilson:1980:a}, and a type of desert leaf cutting ant which lives in colonies that periodically disband and which are founded by unrelated females from a mating swarm \citep{Rissing:1989:a}. However, the issues explored by the general form of the model are broadly applicable to a wide range of population structures where individuals have the majority of their interactions with a subset of the population. An algorithmic description of such an aggregation and dispersal process is as follows:

\begin{enumerate}
	\item \textbf{Initialisation}: Initialise the migrant pool with $N$ individuals;
	\item \textbf{Group formation (aggregation)}: Assign individuals in the migrant pool randomly to groups of size $S$;
	\item \textbf{Reproduction}: Perform reproduction within groups for $t$ time-steps, as described in the next section;
	\item \textbf{Migrant pool formation (dispersal)}: Return the progeny of each group to the migrant pool;
	\item \textbf{Iteration}: Repeat from step 2 onwards for a number of generations, $T$.
\end{enumerate}

\begin{figure}
\centering
\epsfig{file=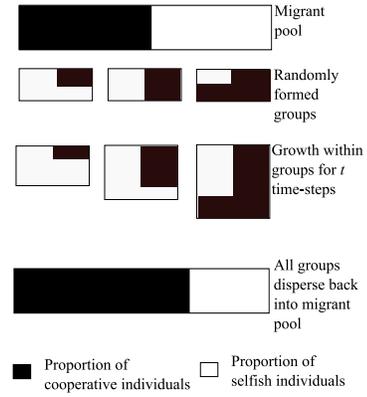, scale = 0.23} 
\caption{Cooperation can increase in frequency in the migrant pool due to differential group contributions, even though it decreases in frequency within each group.}
\label{figAggDisp}
\end{figure}

Group competition occurs in a population structured in this fashion during the dispersal stage, since groups that have grown to a larger size will contribute more individuals to the migrant pool. Crucially, this means that cooperative traits that are individually disadvantageous but that benefit the group can potentially increase in frequency (see Figure~\ref{figAggDisp}), depending on the balance of within- and between-group selection. 

A key factor in determining the balance between the levels of selection in any model of group selection is the variance in group composition \citep{Wilson:1980:a}, for if there is no variance then there is nothing for selection to act on \citep{Darwin:1859:a}. In particular, there must be variance in initial group composition which causes a variance in group size prior to dispersal. In the above model, this variance is generated through random sampling of individuals from the migrant pool, where the sample size corresponds to the initial group size. Since an increased sample size causes a decrease in between-sample variance, increasing the initial group size decreases between-group variance and hence the efficacy of group selection \citep{Wade:1978:a}. It is therefore often assumed that the upper limits of group size which produce a non-zero group selection effect are very small. 

However, in this paper we present results which suggest that this follows from an assumption of within-group dynamics that lead to the competitive exclusion of a cooperative type by its selfish counterpart at within-group equilibrium. Specifically, we are able to show that where within-group dynamics instead lead to a stable coexistence of types, then the range of initial group sizes over which an effect of group selection can be seen is much larger. This is due to the fact that since the cooperative type cannot be driven extinct, variance in group composition when sampling from the migrant pool is always possible. In the next section, we describe how both competitive exclusion and coexistence dynamics can be modelled within groups.

\subsection{Within-group Dynamics: Competitive Exclusion Versus Coexistence}

Classical models of group selection consider a scenario where a selfish type ultimately drives its cooperative counterpart extinct at within-group equilibrium. In particular, fitness functions of the following form, first proposed by \citet{Wright:1945:a} but subsequently used in a plethora of other models \citep{Williams:1957:a,Smith:1964:a,Charnov:1975:a,Wilson:1980:a,Wilson:1987:a}, are typically used to model within-group selection:

\begin{eqnarray}
f_s & = & 1+p_cg \label{eqnFitSelfish}\\
f_c & = & (1+p_cg)(1-a) \label{eqnFitCoop}
\end{eqnarray}

In the above equations, $f_s$ and $f_c$ denote the per capita fitness of selfish and cooperative individuals within a group, respectively. Cooperators, whose proportion within the group is denoted by $p_c$, confer a fitness benefit $g$ on every group member. Crucially, both types receive this benefit, while only cooperators pay a cost, represented by the selection coefficient against cooperation, $a$. It is then clear that if these equations are iterated until equilibrium is reached then the selfish type will be driven to fixation within the group.
 
Both competitive exclusion and stable coexistence within-group dynamics can instead be modelled using the standard two-species symmetric Lotka-Volterra competition equations (e.g. \citep{May:1976:a}). For implementation purposes, we use the following difference equation as a discrete approximation:

\begin{equation}
N_{i(t+1)} = N_{i(t)} + \left[ M_i N_i \left( {\frac{{K_i  - \alpha _{ii} N_i  - \alpha _{ij} N_j }}{{K_i }}} \right)\right].
\label{eqLV}
\end{equation}

In the above equation, $N_{i(t)}$ is the biomass of species $i$ at time $t$, and $M_i$ the intrinsic per capita growth rate. Each species has an intrinsic carrying capacity $K_i$, which is then modified through interspecific interactions. Specifically, the per capita effect of species $j$ on species $i$ is given by $\alpha _{ij}$, the coefficient of interaction. All such interactions are competitive in the above equation, as ensured by the negative sign and the stipulation that all $\alpha > 0$. Similarly, $\alpha _{ii}$ denotes the negative density-dependant effect of species $i$ on itself that prevents unbounded exponential growth. This coefficient can be seen as representing crowding and can vary for different species.

We define selfish ($s$) and cooperative ($c$) strategies in the above equation through settings of the within- and between-type interaction coefficients. Specifically, a selfish type is defined as having a large negative per capita effect on both itself ($\alpha_{ss}$) and the other type ($\alpha_{cs}$). A cooperative type is then defined as having correspondingly smaller per capita negative effects ($\alpha_{cc}$ and $\alpha_{sc}$). A pure group of cooperators will therefore grow to a larger size than a pure group of selfish individuals, creating a group productivity differential on which selection can potentially act. However, within mixed-groups the selfish type will reach the larger frequency (provided that $\alpha_{ss}$ is not too large), since $\alpha_{cs}>\alpha_{sc}$. In other words, cooperators are favoured by between-group selection, while selfish individuals are favoured under within-group selection, exactly as in a classic group selection scenario. It should be noted that our definition of cooperative behaviour corresponds to weak, rather than strong, altruism \citep{Wilson:1980:a}. This follows because although cooperation confers a relative fitness disadvantage compared to a selfish individual within the same group, it nevertheless increases the absolute fitness of all group members, including the cooperator.

It is well known that a stable coexistence of both types occurs in such a model when competition for resources (space, food, etc.) between individuals of the \emph{same} type is stronger than competition between individuals of different types. Such a case corresponds at the ecological level to species occupying different niches, i.e. only partially overlapping in their resource requirements \citep{May:1976:a}. Conversely, if between-type competition is stronger than within-type competition then competitive exclusion of one type will occur. Between- and within-type competition are both modelled in the Lotka-Volterra equations through the settings of the interaction coefficients. Throughout this paper, we assume the following:     

\begin{enumerate}
	\item $\alpha_{cc}<\alpha_{ss}$ and $\alpha_{sc}<\alpha_{cs}$, i.e. that cooperators have lower negative density-dependant effects on themselves and others;
	\item $\alpha_{cs}\alpha_{sc} \le 1$;
	\item $M$ and $K$, the intrinsic per capita growth rates and carrying capacities respectively, are the same for both types.
\end{enumerate}

Given these assumptions, competitive exclusion of the cooperative type occurs when $\alpha_{ss}<\alpha_{cs}$, producing qualitatively similar dynamics to those of the traditional within-group selection equations (\ref{eqnFitSelfish}) and (\ref{eqnFitCoop}). However, when $\alpha_{ss}>\alpha_{cs}$ then the cooperative type is maintained at within-group equilibrium at an above-zero frequency, i.e. a stable coexistence of types occurs. When the interaction coefficients of the cooperative strategy are fixed, the equilibrium frequency at which it is maintained then depends upon the settings of $\alpha_{ss}$ and $\alpha_{cs}$, i.e. the magnitude of the negative density-dependant effects of selfish individuals on themselves and cooperators, respectively. In addition, the within-group equilibrium is reached more quickly the greater the effects of the selfish type. Although a Lotka-Volterra model such as this is typically interpreted at the ecological level as representing species interactions, it could also be interpreted as a model of allelic competition dynamics within a single species group. In particular, coexistence Lotka-Volterra dynamics are analogous to balancing selection for a stable allelic polymorphism within a group. Conversely, competitive exclusion of one species by another is analogous to directional selection driving one allele to fixation. The motivation for using the language of allelic competition is to facilitate comparison with classical group selection models, which consider competition between selfish and cooperative alleles.

Our use of the Lotka-Volterra equations in this paper should be contrasted from their use in community or ecosystem selection models \citep{Wilson:1992:a,Penn:2003:a}. Such models do not consider explicit cooperative and selfish types in the fashion of traditional group selection models. Instead, they examine the complex within-group dynamics that arise when a larger number of types are present. These complex dynamics can give rise to multiple within-group attractors, which can then provide a source of variation in their own right upon which selection can act \citep{Penn:2003:a}. By contrast, in this paper we consider simple two-type within-group dynamics, where only a single group attractor exists (either coexistence or competitive exclusion, as discussed above). As far as we are aware, our use of the Lotka-Volterra equations to define explicit selfish and cooperative strategies is novel.

\section{Results}

\begin{table*}[!t]
	\centering
	
		\begin{tabular}{|l|l|l|}
		\hline
		\textbf{Parameter} & \textbf{Value (competitive exclusion)} & \textbf{Value (coexistance)} \\ \hline
		\textbf{$\alpha_{ss}$} & \textbf{1.9} & \textbf{2}\\ \hline
		\textbf{$\alpha_{cs}$} & \textbf{2} & \textbf{1.9}\\ \hline
		$\alpha_{cc}$ & 1 & 1\\ \hline
		$\alpha_{sc}$ & 0.5 & 0.5\\ \hline
		$K$ & 100 & 100\\ \hline
		$M$ & 0.1 & 0.1 \\ \hline 
		\end{tabular}
	\caption{Parameter settings of the Lotka-Volterra equation. Note that the only difference between the competitive exclusion and coexistence settings is a swapping of the values of $\alpha_{ss}$ and $\alpha_{cs}$.}
	\label{tabParams}
\end{table*}

The parameter settings used for the Lotka-Volterra equations throughout this paper are shown in Table~\ref{tabParams}. Changing between competitive exclusion and coexistence within-group dynamics is achieved by simply switching over the values of $\alpha_{ss}$ and $\alpha_{cs}$, since that determines whether $\alpha_{ss}<\alpha_{cs}$ and hence whether competitive exclusion occurs. The values of the interaction coefficients in Table~\ref{tabParams} produce representative within- and between-group dynamics; other settings produce the same qualitative trends. In this section, we first present results using classical competitive exclusion dynamics, and then contrast these to results from the coexistence case.

\subsection{Group Selection Dynamics in the Competitive Exclusion Case}
\begin{figure}[ht]

\centering
\subfigure[Biomass of each type.]{
\epsfig{file=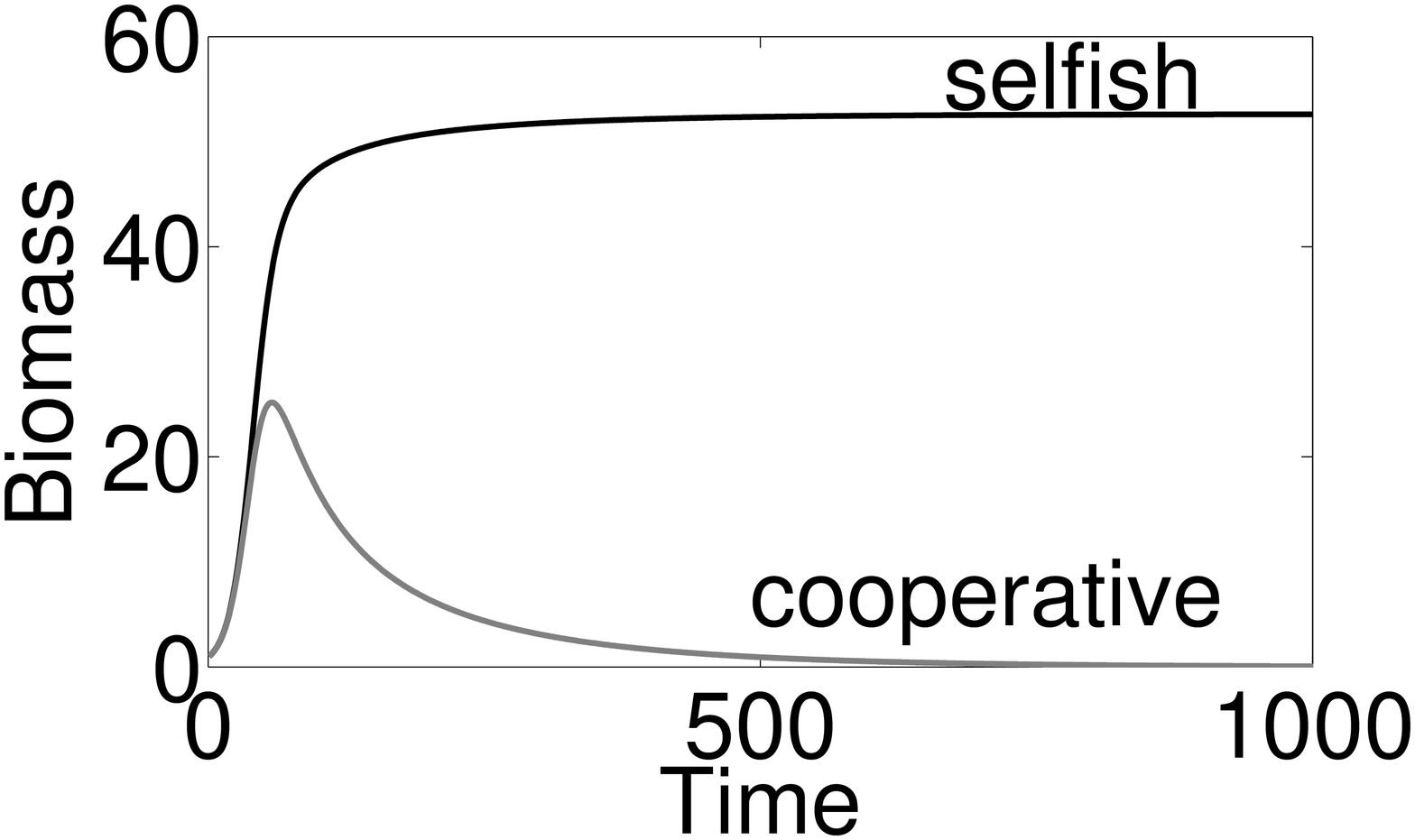, scale = 0.14}
\label{figCompExcBiomass}
}
\subfigure[Proportion of selfish type.]{
\epsfig{file=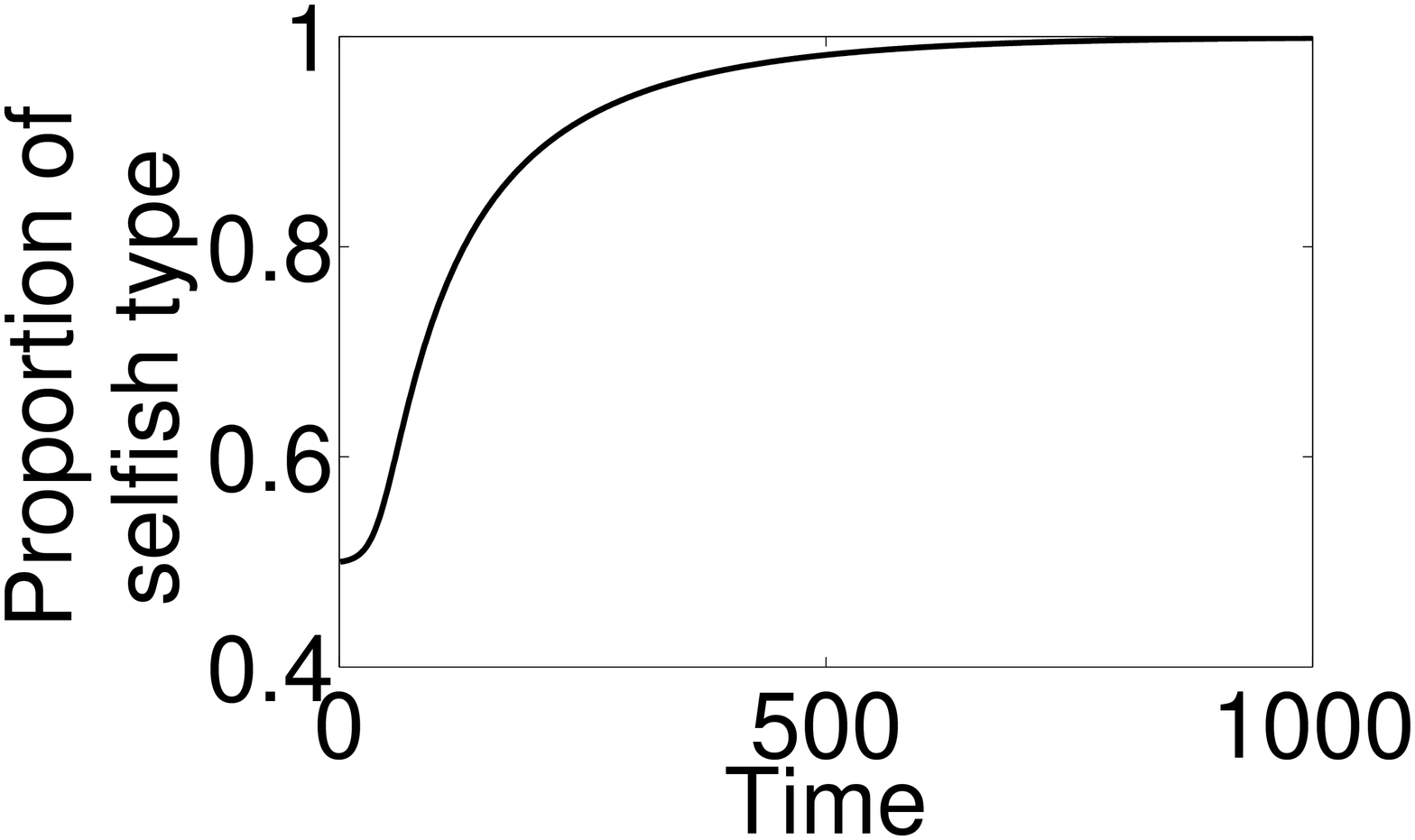, scale = 0.14}
\label{figCompExcProp}
}
\caption{Competitive exclusion within-group dynamics.}
\end{figure}

The within-group dynamics for a group initialised with unit biomass of each type are shown in Figures~\ref{figCompExcBiomass} and \ref{figCompExcProp}, for the competitive exclusion case. Initially, both types are in their growth phase; their biomass is below the intrinsic type carrying capacity of 100. However, the selfish type grows at a faster rate, despite the fact that their intrinsic growth rates, $M$, are the same. This is because of the greater negative density-dependant effect of the selfish type on cooperators, i.e. $\alpha_{cs}>\alpha_{sc}$. Finally, since $\alpha_{cs}>\alpha_{ss}$, the cooperative type is driven to extinction. Furthermore, as Figure~\ref{figCompExcProp} shows, the proportion of selfish individuals increases monotonically. Such behaviour is qualitatively identical to that of directional within-group selection for a selfish allele in classical group selection models (e.g. \citep{Wright:1945:a,Wilson:1980:a}).

\begin{figure}
\centering
\epsfig{file=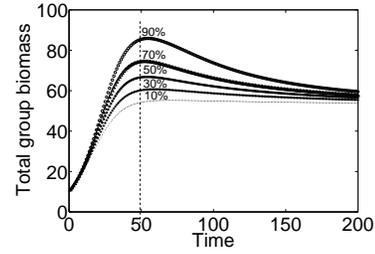, scale = 0.15} 
\caption{Final group size as a function of time spent reproducing within groups; initial group size 10 with various \% of cooperators (competitive exclusion case shown; coexistence quantitatively similar). Dotted line shows time at which difference in group productivity is greatest.}
\label{figDiffProd}
\end{figure}

Now let us consider global dynamics under group selection in this competitive exclusion case. In order for group selection to operate through an aggregation and dispersal process, a difference in group size at the dispersal stage must exist. Figure~\ref{figDiffProd} illustrates how final group size varies as a function of the time spent in the group prior to dispersal, for various starting frequencies of cooperators in groups of initial size 10. It can be seen from this graph that, using the parameters described in Table~\ref{tabParams}, groups with a greater proportion of cooperators do indeed grow to a larger size. In addition, the results for the coexistence case, where $\alpha_{ss}$ and $\alpha_{cs}$ are swapped, are quantitatively similar. These results therefore confirm that group selection can in principal operate, since there is a variation in group productivity on which selection can act.

\begin{figure}
\centering
\epsfig{file=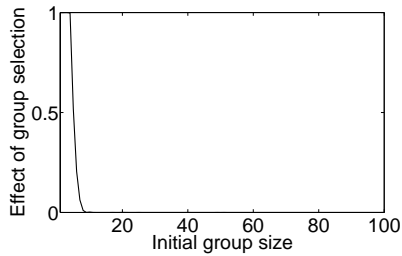, scale = 0.205} 
\caption{`Effect of group selection' (see text) as a function of initial group size in the competitive exclusion case.}
\label{figCompExcVarySize}
\end{figure}

To determine the magnitude of the effect of group selection, the aggregation and dispersal process was executed for 5000 iterations, which preliminary experimentation had shown to be a sufficient length of time for a global equilibrium to be reached, using the within-group parameter settings described in Table~\ref{tabParams}. Equation~\ref{eqLV} was iterated 30 times in the reproduction stage, while the global population size was maintained at 5000. Initial group size was then varied from 1 to 100 inclusive, while the migrant pool was initialised with 50\% of each type. The result of this process after 5000 aggregation and dispersal cycles is shown in Figure~\ref{figCompExcVarySize}, where `effect of group selection' on the $y$-axis is defined as the difference between the frequency of the selfish type at within-group equilibrium and the global frequency of the selfish type after 5000 aggregation and dispersal cycles. Since the within-group equilibrium in the competitive exclusion case is the selfish type at 100\%, the $y$-axis equivilantely shows the global frequency of cooperators in this case. Furthermore, it should also be stressed that the within-group equilibrium is the equilibrium that would be reached in an unstructured population where there were no groups. The $y$-axis therefore shows the effect that group structure is having on the outcome of evolution compared to that in an unstructured population.

There are two points to note from Figure~\ref{figCompExcVarySize}. Firstly, increasing the initial group size decreases the effect of group selection, and consequently the global proportion of cooperators. In particular, for small group sizes, the cooperative type reaches global fixation (and remains there because we do not reintroduce types by mutation). However, for group sizes above 10, it is driven extinct. This follows from the fact that the between-group variance necessary for group selection to act is generated by random sampling from the migrant pool, and therefore rests on the existence of a small initial group size, as previously discussed.

The second, and a key point for this paper, is that the effect of group selection rapidly tends to zero as initial group size increases. Specifically, above a group size of 10, there is no measurable effect at all. Such a result may therefore make the idea of group selection acting on randomly formed groups seem rather implausible as a significant evolutionary pathway. However, the above results only consider the competitive exclusion case; in the coexistence case, the results are somewhat different, as shown in the following section.

\subsection{The Efficacy of Group Selection under Coexistence Dynamics}

\begin{figure}[h]
\subfigure[Biomass of each type.]{
\epsfig{file=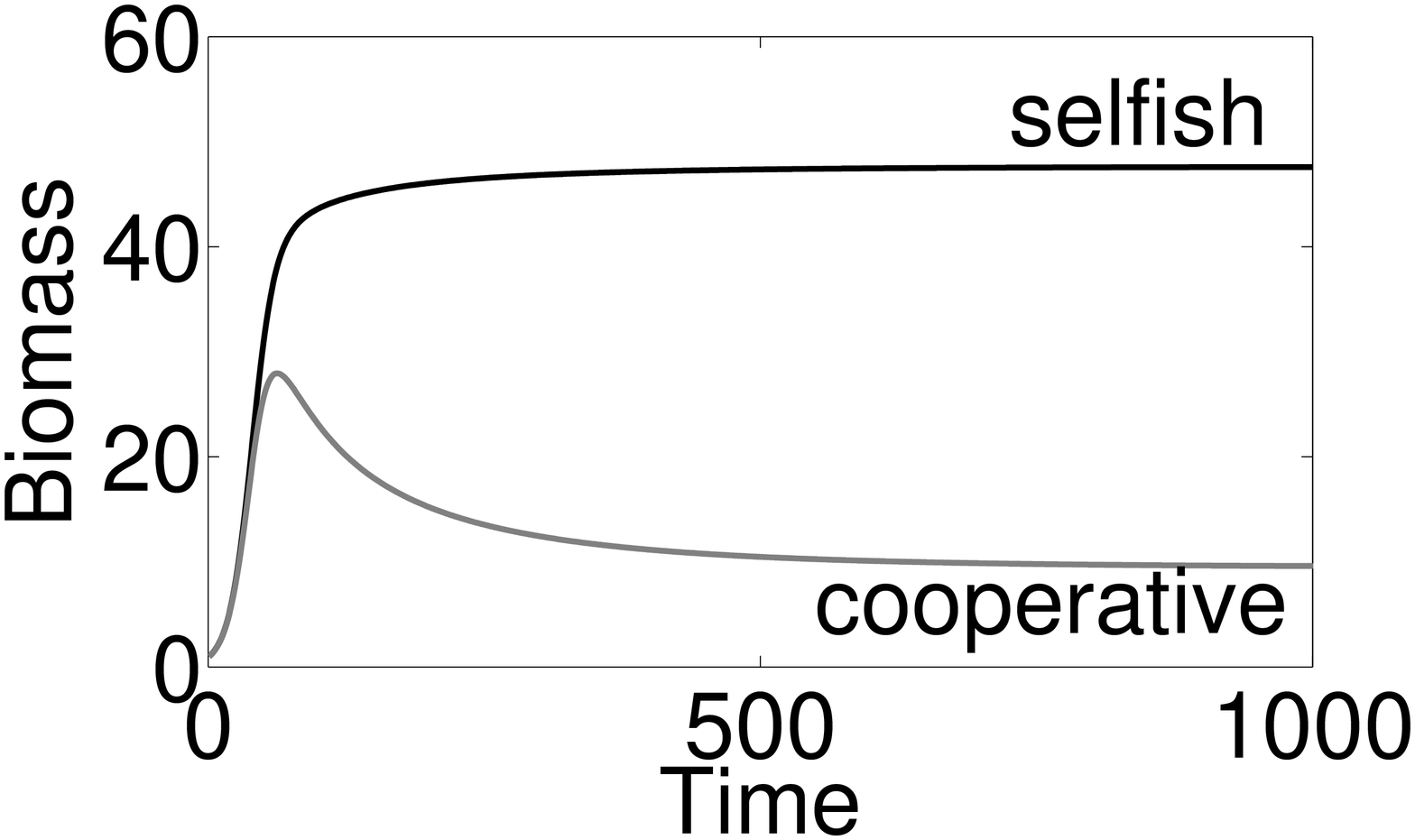, scale = 0.14}
\label{figCoexistenceBiomass}
}
\subfigure[Proportion of selfish type.]{
\epsfig{file=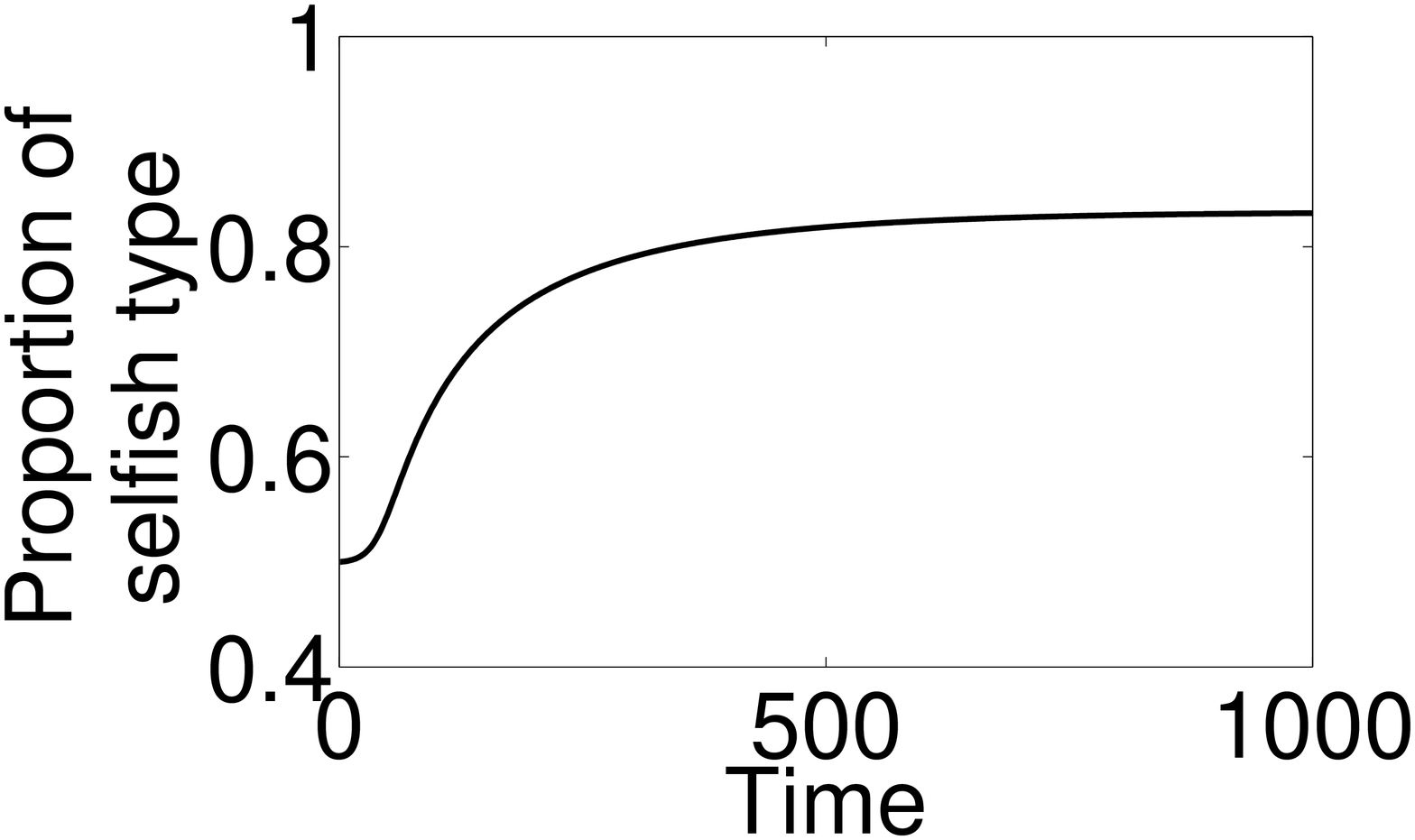, scale = 0.14}
\label{figCoexistenceProp}
}
\caption{Coexistence within-group dynamics.}
\end{figure}

Let us now consider the coexistence dynamics that arise from redefining the selfish strategy as $\alpha_{ss}=2$ and $\alpha_{cs}=1.9$, i.e. by swapping the interaction coefficients over from the competitive exclusion case. Figures~\ref{figCoexistenceBiomass} and \ref{figCoexistenceProp} show how the cooperative type is no longer driven extinct at the within-group equilibrium. In particular, the change in the frequency of the selfish type from an initialisation of 50\% shows clear balancing selective dynamics resulting in the maintance of cooperation at an above-zero frequency. In other words, the result is a stable coexistence of cooperative and selfish types within a group.

\begin{figure}
\centering
\epsfig{file=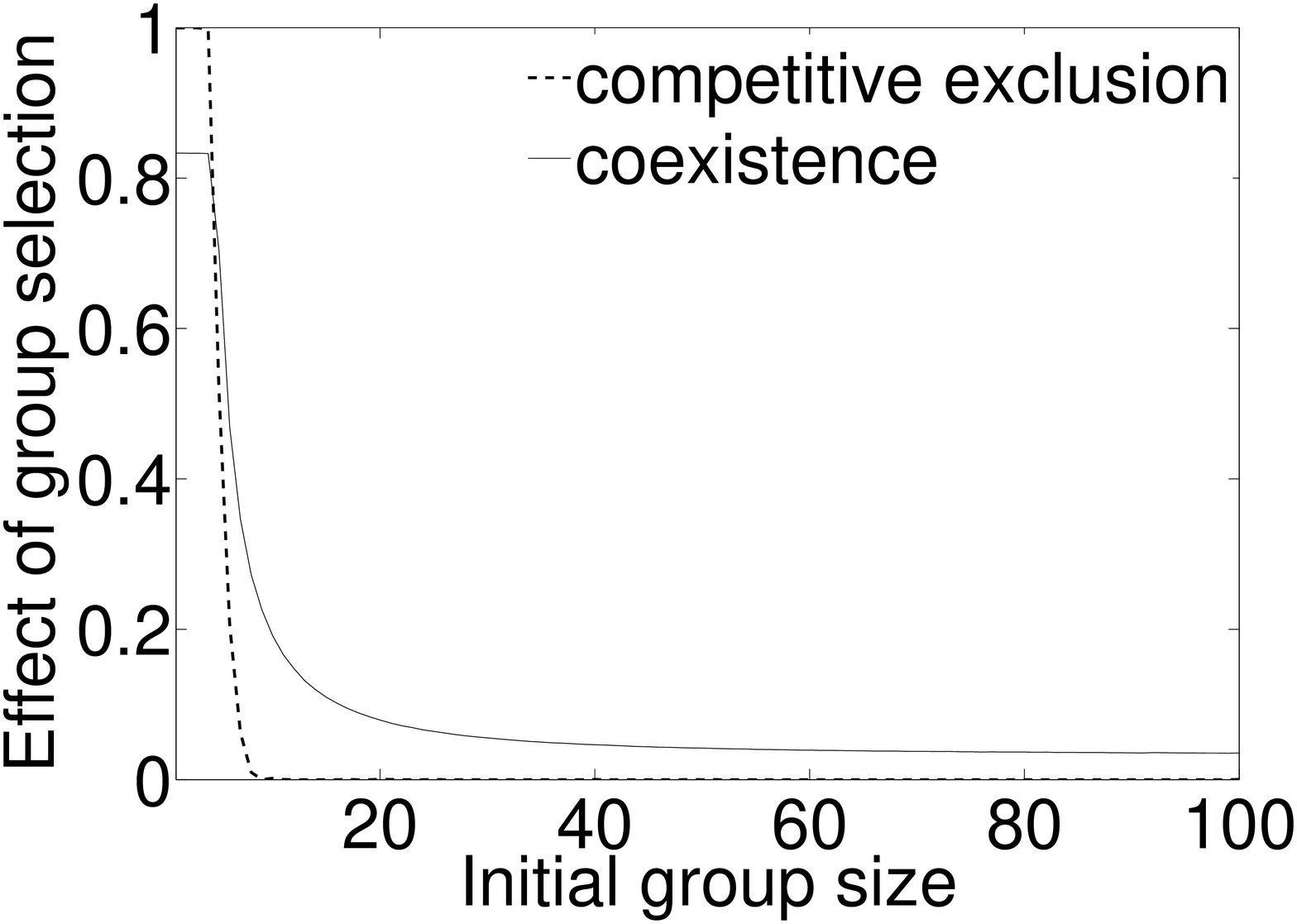, scale = 0.2} 
\caption{Comparing the range of group sizes over which an `effect of group selection' (see text) can be seen between coexistence and competitive exclusion dynamics. }
\label{figBothVarySize}
\end{figure}

Group selection dynamics under the aggregation and dispersal process are now as shown in the black curve in Figure~\ref{figBothVarySize}. Crucially, in contrast to the competitive exclusion case (shown again in the dotted line), an effect of group selection can be seen over the entire range of group sizes examined. For example, in groups of initial size 50, group selection can be seen to still increase the global frequency of cooperation above the within-group equilibrium. The significance of this observation is that since the within-group equilibrium is the same equilibrium that would be reached in an unstructured population, these results show that group structure is having an effect on population dynamics across a wide range of group sizes. 

\begin{figure}
\centering
\epsfig{file=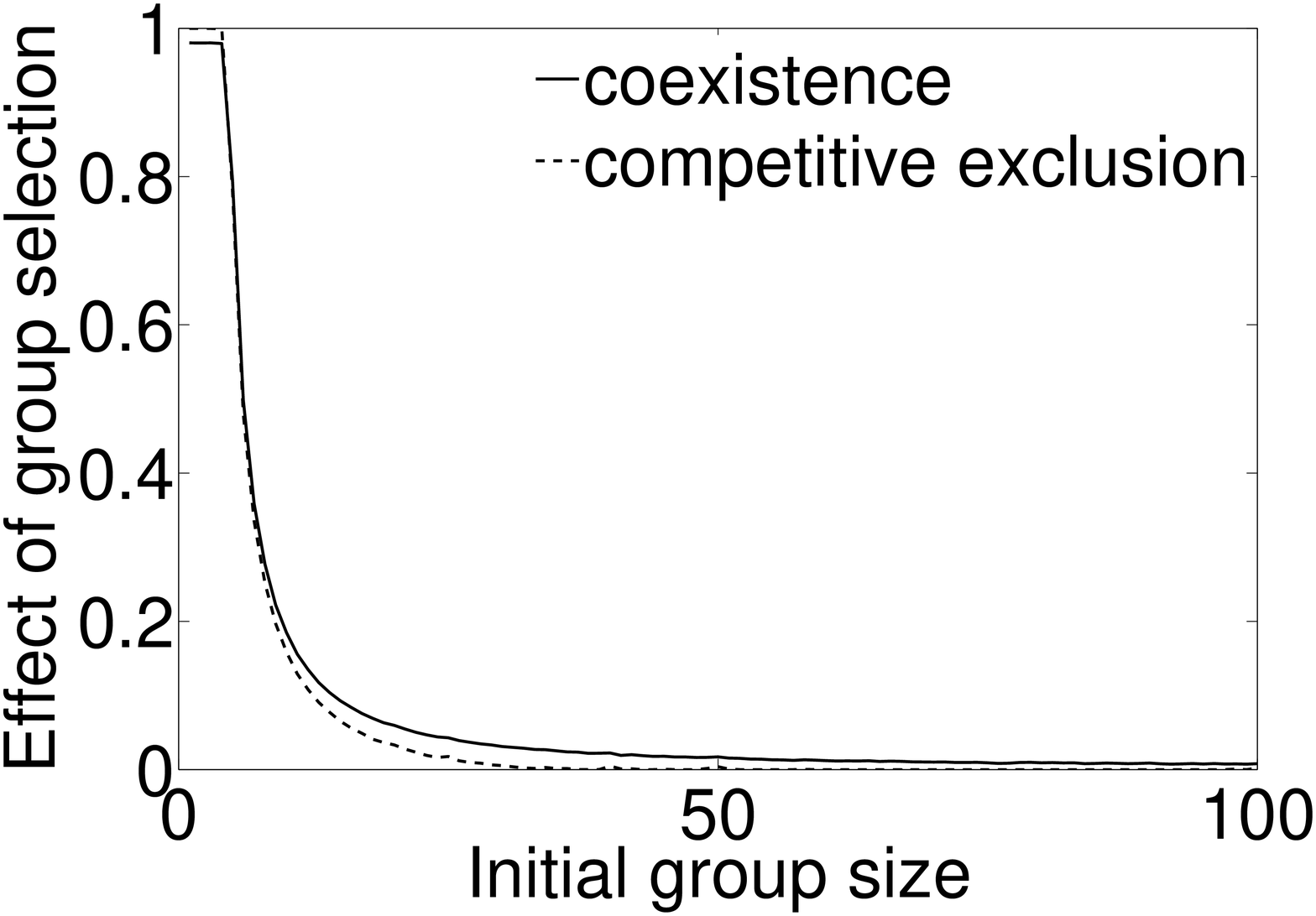, scale = 0.21} 
\caption{Demonstrating that the same qualitative trends arise where within-group selection towards selfish behaviour is stronger in the coexistence case than in Figure~\ref{figBothVarySize}. Here, $\alpha_{cs}=1.99$ and $\alpha_{ss}=2$ in the coexistence case, vice versa for competitive exclusion; all other parameters as in Table~\ref{tabParams}.}  
\label{fig199}
\end{figure}

Finally, to verify that this result is not an artefact of the particular values of $\alpha_{ss}$ and $\alpha_{cs}$ used, the same curves were plotted for a variety of other parameters. Figure~\ref{fig199} provides an example of this, where $\alpha_{ss}=1.99$ and $\alpha_{cs}=2$ in the competitive exclusion case, vice versa for the coexistence case. These parameters were chosen since they represent stronger within-group selection towards selfish behaviour in the coexistence case than in the previous example. Specifically, the within-group equilibrium frequency of the selfish type in the coexistence case is 98.04\%, compared to 83.3\% previously. The results in Figure~\ref{fig199} show that while an effect of group selection is still seen over a larger range of group sizes in the coexistence case, the magnitude of the effect is reduced compared to Figure~\ref{figBothVarySize}. The reason for this is that variance in group composition is proportional to the frequency of cooperators in the migrant pool in this case, and hence to the corresponding within-group equilibrium frequency, as discussed in detail in the following section. 

\section{Discussion}

The results in the previous section demonstrate that where a stable coexistence of types occurs at within-group equilibrium, an effect of group selection on global frequencies can be seen over a much larger range of initial group sizes than in the competitive exclusion case. In particular, as group size increases in the competitive exclusion case, any measurable effect of group selection on the global frequency of cooperation rapidly tends to zero. By contrast, in the coexistence case, some effect on global frequencies is seen over the entire range of group sizes examined. It must be stressed that we do not make a particular claim about the magnitude of the effect for large group sizes. Rather, our model implies that there is \emph{some} measurable effect on frequencies over a large range of group sizes; how large this effect may be will depend on the properties of the natural system under consideration. However, the fact that any effect of group selection still exists over a large range of parameters is significant, since it suggests that where within-group dynamics in nature are of the coexistence type, some effect of a group population structure may always be acting. 

Coexistence dynamics allow group selection effects to be sustained over a larger range of group sizes because of the effect of migrant pool frequencies on between-group variance. In particular, because group formation constitutes random sampling from the migrant pool, initial between-group variance can be approximated by the binomial distribution, and is then given by $p_cp_s/S$, where $p_c$ is the proportion of the cooperative type in the migrant pool, $p_s$ the proportion of the selfish type, and S the initial group size \citep[p. 27]{Wilson:1980:a}. Since $p_c+p_s=1$, it follows that between-group variance is proportional to the frequency of the least frequent type, i.e. variance is maximal when both types are of equal frequency, and zero when one type is at fixation. Therefore, where one type reaches global fixation then there can be no variance and hence no group selection. However, in the coexistence case, where one type cannot reach fixation, it follows that there must always be some variance and hence some possible effect of group selection. The fact that the variance is proportional to the frequency of the least frequent type is illustrated by the difference between Figures~\ref{figBothVarySize} and \ref{fig199}, where the lower within-group equilibrium frequency of cooperators in Figure~\ref{fig199} results in a reduced effect of group selection for large group sizes.

A further observation from Figure~\ref{figBothVarySize} is that a gradient towards an increased effect of group selection also exists over a larger range of group sizes in the coexistence case. Specifically, decreasing group size by a small amount yields an increase in the effect of group selection for groups of size 20 in the coexistence case. However, there is no gradient at this size in the competitive exclusion case. The significance of this is that increasing the effect of group selection increases average absolute individual fitness in the population, due to an increased global level of cooperation. If group size can be partly determined by individual traits \citep{Powers:2007:b} then this may provide an adaptive gradient towards smaller groups, increased levels of cooperation, and greater fitness. In the competitive exclusion case, however, such a gradient only exists over a much smaller range of group sizes. While numerical experimentation investigating whether either gradient can be followed by a series of small mutations will be the subject of a future study, the results presented here do suggest that the concurrent evolution of group size and cooperation is more plausible in cases where a stable coexistence of types within groups exists.

\section{Conclusions}

Any group selection process requires there to be a variation in group composition. In aggregation and dispersal style models, this variation arises through the random assignment of individuals from the migrant pool into groups. Consequently, it is often suggested that an effect of group selection on the global frequency of types will only be seen for very small initial group sizes. However, the models on which this claim is based typically only consider within-group dynamics that lead to the competitive exclusion of a cooperative type by its selfish counterpart.

In this paper, we suggest that within-group competitive exclusion dynamics may be an unnecessary assumption in a number of situations, including the modelling of multi-species consortia in bacterial biofilms and during egalitarian major transitions. A model has been presented which shows that where such coexistence dynamics are present within groups, the range of initial group sizes over which an effect of group selection can be seen is much larger. Consequently, the potential for an adaptive gradient towards smaller groups and increased cooperation exists over a much larger range of group sizes under coexistence dynamics. This is in sharp contrast to the competitive exclusion case, where the effects of group selection rapidly reach zero as initial group size increases, excluding the possibility of such a gradient for a large range of parameters. Our results suggest that, where group size can be influenced by individual traits, the evolution of smaller groups and increased cooperation is more plausible under coexistence dynamics. Such increased group cooperation is a vital component of many major transitions in evolution \citep{Smith:1995:a,Michod:1999:a}.

We have shown in this paper that the conventional conclusion that group selection effects can only be seen for very small groups rests on the assumption that within-group dynamics lead to competitive exclusion. If a within-group coexistence of competing types is instead permitted, then the range of group sizes over which an effect can be seen is much larger. This result follows from the fact that the variance in group composition upon which group selection acts is dependant not only on group size but also on the frequencies of types in the migrant pool. In particular, since neither type can be driven extinct under coexistence dynamics, there will always be some variance in group composition when sampling from the migrant pool, which can then be acted on by group selection. Thus, since it is not necessary to assume that within-group dynamics lead to competitive exclusion, this result shows that group selection can operate in a wider range of conditions than previously realised.

\footnotesize  
\bibliographystyle{apalike} 
\bibliography{lit}

\end{document}